*Applications of Blockchain for the Governance of Integrated Project Delivery: A Crypto Commons Approach*

Jens J. Hunhevicz[1*], Pierre-Antoine Brasey[1,2], Marcella M. M. Bonanomi[1,3], Daniel M. Hall[1], Martin Fischer[4]

[1] ETH Zurich, Institute of Construction and Infrastructure Management, Chair of Innovative and Industrial Construction, Zurich, Switzerland

[2] INGPHI Ltd, Concepteurs d'ouvrages d'art, Lausanne, Switzerland

[3] Polis Lombardia, Milan, Italy

[4] Stanford University, Department of Civil and Environmental Engineering, Stanford, United States

[*] Corresponding author: hunhevicz@ibi.baug.ethz.ch

**Abstract**

This paper outlines why and how blockchain can digitally support and evolve the governance of collaborative project deliveries, such as integrated project deliveries (IPDs), to provide the foundation for novel and disruptive forms of organizational collaboration in the construction industry. Previous work has conceptualized IPDs as a common pool resource (CPR) scenario, where shared resources are collectively governed. Through the use of blockchain and smart contracts for trustworthy peer-to-peer transactions and execution logic, Ostrom's design principles can be digitally encoded to scale CPR scenarios. Building on the identified connections, the paper 1) synthesizes fourteen blockchain-based mechanisms to govern CPRs, 2) identifies twenty-two applications of these mechanisms to govern IPDs, and 3) introduces a conceptualization of the above relationships towards a holistic understanding of collaborative project deliveries on the crypto commons for novel collective organization of construction project delivery between both humans and machines.

*Keywords: Integrated Project Delivery (IPD), Common Pool Resource (CPR), Ostrom Principles, Blockchain, Distributed Ledger Technology (DLT), Smart Contracts, Decentralized Autonomous Organization (DAO), Construction Automation*

## 1. Introduction

Construction project delivery models (PDMs) describe how the multiple parties involved in a project are organized and managed to create and capture value (Davies et al. 2019). Even though the construction industry has been slow in adopting digitalization, new digital technologies and processes slowly make their way into the construction industry (Singh 2019). Digital information is changing how projects are delivered (Whyte 2019); it can motivate the development of novel collaborative PDMs with new incentive structures, procurement methods, and approaches to communication.

Meanwhile, digital information technologies in the construction industry are also rapidly changing. One technology that is increasingly researched for the construction industry is blockchain (Li et al. 2019; Li and Kassem 2021; Nawari and Ravindran 2019a; Perera et al. 2020; Wang et al. 2017). Blockchain is a particular design option of distributed ledger technology (DLT) (Ballandies et al. 2021; Tasca and Tessone 2019) that enables direct peer-to-peer transactions of value without relying on trusted facilitators.

The first ever blockchain created is Bitcoin (Nakamoto 2008). Since then, many new blockchains iterated on the approach of Bitcoin to enable new features and infrastructure (Spychiger et al. 2021a). Most notable, the Ethereum blockchain (Buterin 2014) made it possible that Turing-complete code pieces termed *smart contracts* could be executed on a blockchain. Smart contracts allow for the coding of interaction rules with blockchain transactions for digital workflows to coordinate economic activity of actors in a decentralized and borderless way. In addition, smart contracts can encode containers of value, so-called tokens, such as currencies, securities, or utilities (Ballandies et al. 2021; Mougayar 2017). Tokens can then be transferred among blockchain users.

Blockchain has been repeatedly theorized as promising to improve construction project management practices (Sonmez et al. 2021), especially to support financial management, automatic contract administration, and tracing and securing data along the supply chain (Hewavitharana et al. 2019; Kim et al. 2020). This also aligns well with the most often explored use cases for the construction sector (Hunhevicz and Hall 2020; Li et al. 2019; Li and Kassem 2021; Perera et al. 2020; Scott et al. 2021).

However, scholarship also suggests that the impact of blockchain is highly disruptive to the coordination of existing economic systems (Davidson et al. 2018; Miscione et al. 2019). Smart contracts can create new organizational systems, incentivizing individual actors towards intended collective behaviour (Voshmgir and Zargham 2019). Therefore, blockchain can be an opportunity for new organizational designs governing the upcoming digital reality of PDMs (Hunhevicz et al. 2022a; Sreckovic and Windsperger 2020). Since construction PDMs are already transforming due to increasing digital information (Whyte 2019), there is need to investigate further the impact of the potentially disruptive impact of blockchain.

In this paper, we conceptualize why and how blockchain can digitally support and evolve the governance of PDMs. To build this conceptualization, we specifically make use of ideas about governance of common pool resource (CPR) scenarios (i.e., the "commons") (Ostrom 2015a). Scholars argue that blockchain-based mechanisms can scale CPR scenario governance (Fritsch et al. 2021; Rozas et al. 2021b; a). Such "crypto commons" (Maples 2018) build digital governance structures for commons by leveraging blockchain-based market mechanisms and economic incentives to reward contributions to the common good (Crypto Commons Association 2021).

This is interesting because there is strong theoretical alignment between Integrated Project Delivery (IPD) and the management of CPR scenarios (Hall and Bonanomi 2021). IPD is a new collaborative PDM that uses a relational contracting approach to manage large and complex construction projects. One driver for the development of IPD was the need for more flexible and collaborative organizational structures to gain benefit from digital building information modelling (Hall and Scott 2019). To do this, IPD uses a financial pool to share risk and reward among project participants depending on the outcome of the project. IPD also emphasizes decentralized, agile, and self-organized project governance arranged by the project participants. Collaborative PDMs such as IPD can better deal with the complexity and ever-changing nature of modern construction projects (Levitt 2011; Luo et al. 2017).

The strong alignment of the collective nature of blockchain and collaborative approach of IPD has not escaped the attention of researchers. Nawari and Ravindran (2019b) theorize blockchain as a "evidence of trust" for IPD. Elghaish et al. (2020) and Rahimian et al. (2021) have developed a blockchain prototype for the IPD financial risk-and-reward system. However, these works mainly apply blockchain to improve existing financial processes. As stated above, blockchain has the potential to lead to new forms of organization and governance (Davidson et al. 2018; Jacobo-Romero and Freitas 2021; Miscione et al. 2019), but no work yet has explained how this might occur for construction PDMs.

Therefore, this paper now explores how the relationships of blockchain, CPR theory, and IPD can be used as a theoretical foundation to inform which specific blockchain applications can be developed to evolve and redesign PDMs. This is achieved through systematically exploring the connections between blockchain, CPRs, and IPD. The results of this work can help to conceive the opportunity of blockchain for IPDs to evolve, or even enable the formulation of new digitally supported PDMs on the crypto commons.

## 2. Methodology and Structure of the Paper

An overview of the research approach and contribution is presented in Figure 1. The methodology contained three main steps: 1) We outlined established connections between CPRs, the Ostrom design principles (OPs), and IPDs to manage construction resources; 2) we conducted a state-of-the art review of all papers and articles that propose to use blockchain to manage CPR and identify the proposed mechanisms for the respective OPs; 3) we identified applications of those mechanisms for collaborative construction projects through using the link between IPDs and the OPs. Each of these steps is now described in more detail.

In the departure section 3, we introduce relevant established concepts between CPR theory and IPD that act as basis for our research. First, we introduce Ostrom's design principles (OPs) for the management of CPR scenarios (Section 3.1). Second, we explain the high-level concepts of IPD to manage construction project resources (Section 3.2). Finally, we outline the recently established connection between the OPs and IPD practices (Section 3.3).

To verify the link between blockchain and CPR theory (Section 4), we conducted a comprehensive literature review of all papers and articles that proposed blockchain for the management of CPRs. We identify four journal papers (Fritsch et al. 2021; Pazaitis et al. 2017; Rozas et al. 2021b) and five articles (Dao 2018; Decoodt 2019; Emmett 2019a; de la Rouviere 2018; Schadeck 2019) proposing to govern real-world commons with blockchain-based mechanisms. Building on these works, we cluster and categorize fourteen blockchain governance mechanisms encoding the OPs for the governance of crypto commons.

We then use abductive analysis (Timmermans and Tavory 2012) to theorize how blockchain governance mechanisms can be transferred to the governance of IPDs. *Abduction* is making a probable conclusion from what is known by systematically interpreting, matching, or re-contextualizing phenomena within a contextual framework, from the perspective of a new conceptual framework (Dubois and Gadde 2002b; Kovács and Spens 2005). An abductive approach is fruitful if the objective is to develop the understanding of a "new" phenomenon or new insights about existing phenomena by examining these from a new perspective (Dubois and Gadde 2002b; Kovács and Spens 2005).

To do this, we first synthesized applications based on observed alignment between the blockchain mechanisms for the OPs and IPD practices in line with the OPs (Hall and Bonanomi 2021). We then refined and complemented applications based on supporting blockchain research both from within and outside the construction industry. In total, we identified 22 blockchain applications that can be used to build IPD governance on the crypto commons (Section 5).

A holistic overview of the proposed conceptualization of IPD on the crypto commons demonstrates the cohesiveness between the relationships of the OPs, the blockchain governance mechanisms, and the specific blockchain applications to build novel governance mechanisms for IPDs (Section 6).

The paper ends with a discussion of the opportunities for blockchain to be applied to IPD and other future forms of project delivery, as well as the challenges for governance design and for industry implementations to facilitate next research steps (Section 7).

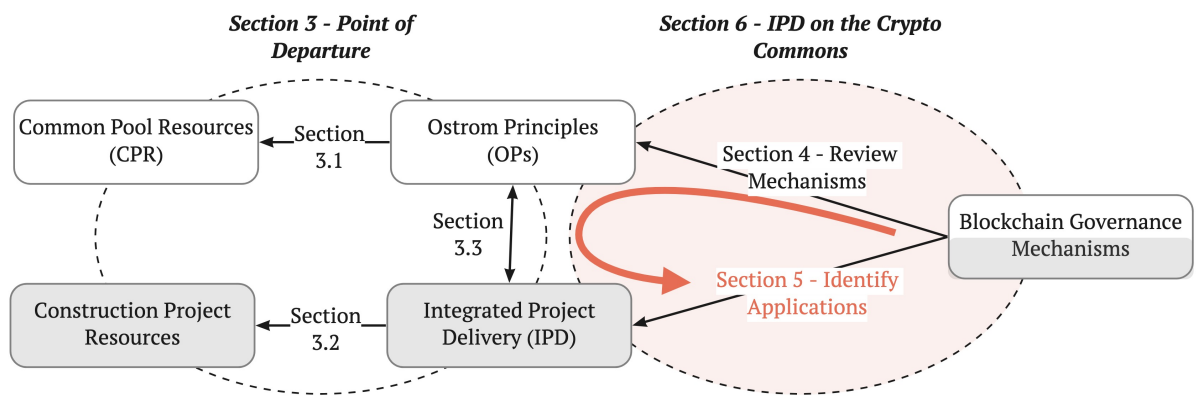

*Figure 1: Schematic representation of the research approach and contribution. CPR related boxes are pictured in white, the IPD related boxes in grey. The paper builds on the existing conceptualization between CPR, OPs, and IPD (see "Point of Departure", Section 3.1 – 3.3. Afterwards, the paper comprehensively reviews literature proposing blockchain for CPRs and the OPs and summarizes the proposed mechanisms (Section 4). Finally, the paper identifies applications of blockchain governance mechanisms for IPD (Section 5) towards a holistic conceptualization of IPD on the crypto commons (Section 6) through abductive reasoning using the connection between the mechanisms, OPs, and IPD practices (see red arrow).*

## 3. Point of Departure

### 3.1. Governing CPR Scenarios

CPRs are natural resources, which are freely shared among many users (Ostrom 1990). Examples include forests, pastures, fishing grounds, parking lots or wiki libraries. In a CPR scenario, users might appropriate resources at a higher than optimal rate, resulting in a downward spiral of total resource availability (Hardin 1968). This is known as the *tragedy of the commons*. For decades, scholars argued that centralized control was the only way to coordinate optimal resource appropriation in CPR scenarios.

However, more recent work pioneered by economist Elinor Ostrom (Ostrom 2010, 2015a; Ostrom et al. 1994) and others (Gardner et al. 1990) overturned these beliefs. Ostrom used case studies to demonstrate that local actors are often successful at self-organizing to better sustain CPR scenarios when compared to centralized interventions. Ostrom identified eight design principles – the OPs – that can guide effective governance of CPR scenarios (Table 1). The OPs explain necessary conditions that should be achieved, to facilitate trust and reciprocity and to sustain collective action in long-lasting CPR scenarios (Cox et al. 2010).

### 3.2. IPD Governance of Construction Projects

IPD is a project delivery model that formally multiple, independent firms to collectively share financial risk and reward among themselves and with the project sponsor during the design and construction of a facility (Lahdenperä 2012). IPD governance today can be best described as the combination of multiple formal and informal practices (Bygballe et al. 2015; Hall and Scott 2019). Such practices include early involvement of key stakeholders, risk and reward mechanisms, joint project control, and target value design (Cheng et al. 2016; Hall et al. 2018).

IPD departs from the traditional model of project delivery in three notable ways (Hall and Bonanomi 2021). First, the multiparty contract of the IPD model creates a shared financial resource pool for the project. The project resources become contractually available for free use by any of the project signatory parties. Second, the participants of IPD projects share decision-making rights over the project governance structures. Decision-making is no longer centralized (Tillmann et al. 2014). Third, the project team shares the financial risks and rewards of the project. Positive outcomes are split among participants. The project teams must self-organize (Bertelsen 2003) and determine who has access to the shared pools and who is allowed to withdraw from this pool.

### *3.3. Governing IPD using CPR Design Principles*

Recent work has proposed a conceptualization bridging the governance of IPD projects and the OPs (Table 1) (Hall and Bonanomi 2021), suggesting that the IPD project environment resembles a CPR scenario (Hall and Bonanomi 2021). Project resources are "pooled" together through a multi-party contract which shares risk and reward (Darrington and Lichtig 2010; Thomsen et al. 2009).

Similar as CPR scenarios must avoid the *tragedy of the commons*, IPD projects must then avoid the *tragedy of the project* – where the project budget and schedule can be subject to over-appropriation by the project stakeholders to the long-term detriment of the project resource system (Hall and Bonanomi 2021). To avoid the tragedy of the project, project managers create effective self-governance structures manifesting in specific management practices for IPDs, which demonstrate many shared characteristics to the OPs. Additional work has validated this connection with examples from IPD project practices (Bonanomi et al. 2019, 2020). Table 1 lists such example practices for IPDs aligned with the OPs for CPRs.

*Table 1: The eight Ostrom principles and their connection with IPD practices (Source: Hall and Bonanomi (2021)).*

| Ostrom Principle (OP) | | Description of OP (Cox et al. 2010; Ostrom 2015b) | Example Practice(s) for IPD (Hall and Bonanomi 2021) |
|---|---|---|---|
| 1 Clearly Defined Boundaries | a) For the users | The boundaries between legitimate and non-users who have right to withdraw resource units from the CPR must be clearly defined. | The participating firms collectively determine who is a risk and reward "partner" and who is not in the multiparty contract. |
| | b) For the resources | Resource boundaries of the system must be clearly defined and separated from the larger socio-economic system. | The project sponsor and project team collectively define which specific aspects of project scope and budget are open to all and which are not. |
| 2 Ensure congruence | a) With local conditions | CPR scenarios should ensure congruence with local conditions of appropriation rules restricting time, place, technology, and/or quantity of resource units. | Trade contractors are engaged early in the project, because they have knowledge of local conditions, such as availability of labor, material, work routines, and other resources. |
| | b) Between appropriation & provision rules | The benefits obtained by users from a CPR, as determined by appropriation rules, should be proportional to the amount of inputs required in the form of labor, material, or money, as determined by provision rules. | The level of participation in the risk/reward pool is weighted according to a firm's individual cost structure or accounting practices, its period of involvement, and/or influence on the outcomes. |
| 3 Collective-choice arrangements | | Most individuals affected by the operational rules can participate in modifying the operational rules. | Firms that have signed the multiparty contract are entitled to participate in management group functions and to vote on decisions that directly concern their work and area of expertise. |
| 4 Monitoring of the users and the resources | a) Presence | Monitors are present to actively audit CPR conditions and appropriator behavior of the users to ensure that all parties are adhering to agreed-upon tasks. | Participants share information on resources, costs, profit, and performance openly and transparently. Teams also create cost targets and then track the weekly withdrawals of resource units, monitoring for deviations (Target Value Design). |
| | b) Accountability | Monitors are accountable to or are the appropriators. | Participants make commitments about the work to be completed. The Planned Percent Completed (PPC) metric tracks the percentage of items promised last week that were completed and is publicly reported to all team members. |
| 5 Graduated sanctions | | Appropriators who violate operational rules are assessed graduated sanctions (depending on the seriousness and context of the offense) by other appropriators, officials accountable to these appropriators, or both. | Sanctions can increase due to continuous non-conformance or underperformance of PPC, leading to the removal of individual participants and/or firms if necessary. |
| 6 Conflict-resolution mechanisms | | Appropriators and officials have rapid access to low-cost local arenas to resolve conflicts among appropriators or between appropriators and officials. | Project participants craft conflict resolution mechanisms that include clear dispute resolution strategies intended to avoid costly litigation proceedings. |
| 7 Minimal recognition of rights to organize | | The rights of appropriators to devise their own institutions are not challenged by external governmental authorities. | Conflict resolution mechanisms allow participants to make collective decisions, including procedures for the team to override the wishes of the project sponsor. |
| 8 Nested enterprises | | Appropriation, provision, monitoring, enforcement, conflict resolution, and governance activities are organized in multiple layers of nested enterprises. | Governance activities of IPD projects are organized into multiple layers of hierarchy using a nested enterprise design. |

# 4. Blockchain and the Crypto Commons: a Review

## 4.1. Blockchain as an Institutional Innovation

The dominant narrative for economic coordination through the blockchain argues that blockchain enables increased productivity of existing processes by lowering transaction costs through costless verification and without the need for costly intermediation (Catalini and Gans 2020). However, some scholars argue the true potential of blockchain is the development of new types of institutional organization with the potential to disrupt and substitute existing economic coordination (Davidson et al. 2018; Jacobo-Romero and Freitas 2021; Miscione et al. 2019). Blockchain is a new way to reach consensus about a shared truth without requiring centralized trust (Davidson et al. 2018). The innovation of blockchain is the consensus protocols using cryptoeconomic mechanisms to reward honest parties to reach consensus on network transactions, e.g. in Bitcoin with proof-of-work (Gervais et al. 2016; Nakamoto 2008). Blockchain disintermediates transactions with a new form of organizational design, and as a consequence can lower transaction costs (Davidson et al. 2018).

As a consequence, applications can leverage the innovation of cryptoeconomic mechanisms of blockchains for trust-minimized social coordination to build new forms of economic activity on top of blockchains. Such cryptoeconomic systems can provide an institutional infrastructure that facilitates a wide range of socio-economic interactions to influence participants in their behavior (Voshmgir and Zargham 2019). There is ongoing exploration of what forms of organization and governance can be supported or replaced through blockchain. Within this paper, we focus on blockchain as a possibility to scale CPR scenarios on the crypto commons.

## 4.2. The Connection of Blockchain and CPR Governance

The OPs describe how commons-based communities can create effective bottom-up governance rules (Cox et al. 2010). However, a major limitation is scaling community governance to large and global systems (Ostrom et al. 1999).

Recent scholars point out that blockchains can be assessed through the lens of CPR theory and the OPs. This can enable the creation of effective bottom-up governance rules for decentralized peer production of the network without any centralized coordination (Red 2019; Shackelford and Myers 2016; Werbach 2020). There is growing recognition that the underlying system governance mechanisms are the key to long-term success of blockchain networks (Beck et al. 2018; Machart and Samadi 2020; Red 2019; Werbach 2020). CPR theory and the OPs are a repeatedly mentioned concept to guide the development of blockchain governance (Shackelford and Myers 2016; Werbach 2020).

Fritsch et al. (2021) find now that blockchain and other DLTs can enable scaling of a new generation of commons-oriented economies, both for digital and physical commons. On the one hand, cryptoeconomic mechanisms decrease the cost of information exchange through minimizing opportunism and uncertainty trough transparency and cryptographic enforcement (Machart and Samadi 2020; Schmidt and Wagner 2019). On the other hand, blockchain provides reliable organizational means to equitably produce and distribute resources in accordance with the shared values of productive communities (Fritsch et al. 2021). The transparent decision-making procedures and decentralized cryptoeconomic incentive systems help avoid the tragedy of the commons (Bollier 2015). The idea is to craft blockchain-based governance mechanisms by encoding the OPs (Rozas et al. 2021b; a). Blockchain could create networked governance to scale real-world commons, similar to how the stock market enabled corporations to scale (Maples 2018). Such crypto commons could allow new types of value creation with crypto assets rather than shares of stock, contributors rather than employees, and decentralized collaboration rather than centralized ownership (Maples 2018).

## 4.3. Blockchain Governance Mechanisms for the Commons

As a basis to later investigate potential applications of blockchain mechanisms for IPD, we reviewed blockchain governance mechanisms proposed for CPRs (see also Section 2). Most notably, Rozas et al. (Rozas et al. 2021a) assesses the relationship between blockchain affordances and the eight OPs to support peer production of real-world commons. Rozas et al. (2021b) explore then how those can be applied to scale-up CPR governance of global software commons to address limitations identified by Stern (2011). Even though IPD can be characterized as a real world common, it hardly falls into the same category of global real world commons. Therefore, we clustered proposed mechanisms from all identified articles into 14 high level mechanisms for the eight OPs (Table 2), instead of just relying on the categorization of Rozas et al. (2021a).

*Table 2: Clustered blockchain governance mechanisms based on reviewed literature.*

| Blockchain Governance Mechanisms | OP | Sources |
|---|---|---|
| M1: Identity, ownership, and access rights based on addresses and tokens | 1a | (Dao 2018; Rozas et al. 2021b; a; Schadeck 2019) |
| M2: Tokenization of the resources | 1b | (Decoodt 2019; Emmett 2019a; Fritsch et al. 2021; de la Rouviere 2018) |
| M3: Decentralized markets to match supply and demand of local needs and conditions | 2a | (Schadeck 2019) |
| M4: Formalizing appropriation and provision rules with smart contracts | 2b | (Dao 2018; Rozas et al. 2021b; a) |
| M5: Decentralized proposal and voting platforms | 3 | (Dao 2018; Emmett 2019a; Rozas et al. 2021b; a; Schadeck 2019) |
| M6: Decentralized prediction markets | 3 | (Dao 2018) |
| M7: Transparent record and automation of transactions | 4a | (Emmett 2019a; Rozas et al. 2021b; a; Schadeck 2019) |
| M8: Digital signatures for tamper-proof commitments | 4b | (Dao 2018; Rozas et al. 2021b; a) |
| M9: Decentralized peer-review mechanisms | 4b | (Pazaitis et al. 2017; Rozas et al. 2021b) |
| M10: Reputation tokens | 4b | (Pazaitis et al. 2017; Schadeck 2019) |
| M11: Transparent and self-enforcing sanctions | 5 | (Dao 2018; Emmett 2019a; Rozas et al. 2021b; a; Schadeck 2019) |
| M12: Decentralized jurisdiction systems | 6 | (Dao 2018; Emmett 2019a; Rozas et al. 2021b; a; Schadeck 2019) |
| M13: Ensure decisions are made by affected parties | 7 | (Rozas et al. 2021b; a) |
| M14: Bottom-up interaction among multiple hierarchical levels | 8 | (Dao 2018; Emmett 2019a; Rozas et al. 2021b; a; Schadeck 2019) |

### 4.3.1. OP1 – Clearly Defined Boundaries

#### a) For the Users

According to OP 1a, the boundaries between legitimate and non-users who have right to withdraw resource units from the CPR must be clearly defined (Cox et al. 2010; Ostrom 2015b). The main identified blockchain mechanism for OP 1a is to govern CPR boundaries through **blockchain addresses and tokens to control identity, ownership, and access rights** (Table 2, M1). Blockchain identifies users with a blockchain address, so there is no need to know the human or machine controlling the address. Access rights and ownership can be assigned to addresses either through smart contract logic that defines roles with specific permissions, or through membership or utility tokens that can be transferred between users (Dao 2018; Rozas et al. 2021b; a; Schadeck 2019). While the second allows to trade these rights with other addresses by transferring the token, the address based roles stays with that address until revoked. In both cases, blockchain controlled ownership and access rights can be more easily and granularly defined, propagated, and revoked (Rozas et al. 2021b).

### b) For the Resources

OP 1b states that resource boundaries of the system must be clearly defined and separated from the larger socio-economic system (Cox et al. 2010; Ostrom 2015b). Within the context of CPRs, **tokenization of the resources** (Table 2, M2) can help to achieve clearly defined resource boundaries on the crypto commons. Once resources are tokenized, cryptoeconomic mechanisms through smart contracts can facilitate a wide range of interaction patterns. Tokenization can be in the form of asset-backed currencies or commodity tokens representing the resource, good, or service in the commons (Fritsch et al. 2021). New mechanisms such as *bonding curves* (Balasanov 2018; Titcomb 2019) can incentivize early protectors of CPR scenarios (Decoodt 2019; Emmett 2019a; de la Rouviere 2018). Bonding curves allow investors to buy a resource token by locking up their investment. Investors can later sell back these tokens according to the new price determined by the bonding curve. The bonding curve increases price with issued supply, and therefore rewards early investors. Bonding curves have been proposed for "continuous organizations", where the underlying tokens represent rights to future revenues (Favre 2019). *Augmented* bonding curves introduce additional functionalities to create a more robust system that is less subjective to speculation and manipulation (Titcomb 2019). They act simultaneously as means of funding, liquidity provider and market maker, while the issued tokens represent access or voting rights to the resource (Zargham et al. 2020). Therefore, augmented bonding curves combine access rights through tokens (M1) with the idea of tokenizing the resource. The interplay between the interests of token holders to sell when token price rises and buy as price drops to claim additional governance power over a growing treasury, creates a negative feedback loop that leverages speculative behavior into a continuous source of income for the commons (Fritsch et al. 2021). The Commons Stack implemented such an augmented bonding curve based on research of Zhargam et al. (2020).

### 4.3.2. OP2 – Ensure Congruence

### a) With Local Conditions

OP 2a states that CPR scenarios should ensure congruence with local conditions of appropriation rules restricting time, place, technology, and/or quantity of resource units (Cox et al. 2010; Ostrom 2015b). **Decentralized markets to match supply and demand of local needs and conditions** (Table 2, M3) are proposed as a blockchain mechanism (Schadeck 2019). Smart contracts encode the rules to trade with other actors not controlled by any intermediary, so the community using the decentralized marketplace can benefit from unrestricted mutual trading. At the same time, the market place can be tailored to comply with the formalized appropriation rules.

### b) Between Appropriation & Provision Rules

According to OP 2b, the benefits obtained by users from a CPR, as determined by appropriation rules, should be proportional to the amount of inputs required in the form of labor, material, or money, as determined by provision rules (Cox et al. 2010; Ostrom 2015b). **Formalizing appropriation and provision rules** (Table 2, M4) with smart contracts can make sure these agreements get obeyed (Dao 2018; Rozas et al. 2021b; a). The transparency of rules also promotes an active discussion of the notion of value in the community (Rozas et al. 2021a). The community can then collectively decide which contributions to recognize, as well as suited local appropriation rules (Rozas et al. 2021b).

### 4.3.3. OP3 – Collective Choice Arrangements

OP 3 states that individuals affected by the operational rules can participate in modifying the operational rules (Cox et al. 2010; Ostrom 2015b). Decentralized decision making and voting are often discussed topics to govern blockchain networks and decentralized applications. It is therefore not surprising that smart contract based **decentralized proposal and voting platforms** (Table 2, M5) are suggested to govern real world commons (Dao 2018; Emmett 2019a). Tokens could grant rights for decision making, either to ensure equal power distribution by design (Rozas et al. 2021b; a), or based on the contribution and reputation of parties (Emmett 2019a; Schadeck 2019).

Furthermore, **decentralized prediction markets** (Table 2, M6) are proposed as a way to establish a trusted knowledge base (Dao 2018). Prediction markets were introduced by Hanson (2013) to establish a more representative picture of a future outcome by using a betting platform. The underlying idea is that predictions made by people willing to risk a loss are more likely to be well-informed.

4.3.4. OP4 – Monitoring

   *a) Presence*

*OP* 4a states that monitors should be present to actively audit CPR conditions and appropriator behavior of the users to ensure that all parties are adhering to agreed-upon tasks (Cox et al. 2010; Ostrom 2015b). Blockchain allows a **transparent record and automation of transactions** (Table 2, M7) through smart contracts of user behavior and participation in the commons observable by all community members (Emmett 2019a; Rozas et al. 2021b; a; Schadeck 2019).

   *b) Accountability*

OP 4b states monitors are accountable to or are the appropriators of a CPR (Cox et al. 2010; Ostrom 2015b). Multiple blockchain mechanisms are proposed to help ensure accountability within CPR scenarios. Every blockchain transaction is signed by a valid private key creating **digital signatures for tamperproof commitments** (Table 2, M8). The immutability and censorship resistance of blockchain ensures that decisions and transactions are accountable since all transactions are transparent and verifiable on the blockchain (Dao 2018; Rozas et al. 2021b; a).

In cases were no automatic checking of work and contributions to the commons is possible, **decentralized peer-review mechanisms** (Table 2, M9) facilitated by smart contracts allow to review the status of work or the perceived value of contributions (Pazaitis et al. 2017; Rozas et al. 2021b).

Pazaitis et al. (2017) proposed then **reputation tokens** (Table 2, M10) to represent the perceived value of contributions in the blockchain system. They can be earned by users through complying with the CPR rules and are hence a measure of accountability (Schadeck 2019).

4.3.5. OP5 – Graduated Sanctions

According to OP 5, appropriators who violate operational rules are assessed graduated sanctions depending on the seriousness and context of the offense (Cox et al. 2010; Ostrom 2015b). Blockchain allows for **transparent and self-enforcing sanctions** (Table 2, M11). Sanctions can be made transparent to the whole community (Schadeck 2019), while smart contracts can self-enforce token-based sanctions (Dao 2018; Emmett 2019a; Rozas et al. 2021b; a; Schadeck 2019) through the loss of either financial or reputation tokens (Dao 2018; Emmett 2019a; Schadeck 2019), or a value-decrease of tokens (Schadeck 2019).

4.3.6. OP6 – Conflict Resolution Mechanisms

OP 6 states that appropriators and their officials should have rapid access to low-cost local arenas to resolve conflicts among appropriators or between appropriators and officials (Cox et al. 2010; Ostrom 2015b). Blockchain offers the possibility for faster conflict resolutions with **decentralized jurisdiction systems** (Table 2, M12) (Dao 2018; Emmett 2019a; Rozas et al. 2021b; a). Tokens ensure skin in the game in disputes, as well as incentivize game theoretic proofs (Schadeck 2019). Such protocols must integrate with existing legal and regulatory systems (Emmett 2019a; Schadeck 2019).

### 4.3.7. OP7 – Minimal Recognition of Rights to Organize

OP 7 states that the rights of appropriators to devise their own institutions should not be challenged by external governmental authorities (Cox et al. 2010; Ostrom 2015b). Within crypto commons, smart contract mechanisms were proposed to **ensure decisions are made by affected parties** (Table 2, M13) (Rozas et al. 2021b; a), e.g. local community rules can only be enforced locally.

### 4.3.8. OP8 – Multiple Layers of Nested Enterprises

OP 8 states that the rules for appropriation, provision, monitoring, enforcement, conflict resolution, and governance activities should be organized in multiple layers of nested enterprises (Cox et al. 2010; Ostrom 2015b). Smart contracts can facilitate **coordination across nested enterprises** (Table 2, M14) between various hierarchical levels of participants to realize shared objectives in the best interest of the commons (Dao 2018; Emmett 2019a; Rozas et al. 2021b; a; Schadeck 2019).

## 5. Applications of Blockchain Governance Mechanisms for IPD

Based on the 14 blockchain mechanisms for CPR scenarios (Table 3), we identified 22 potential applications of blockchain mechanisms for IPDs (Table 3) to govern IPDs as a CPR scenario (Hall and Bonanomi 2021). The methodological approach is explained in Section 2.

We discuss here for each of the 22 identified applications the potential to improve or extend the IPD practices, either based on already existing practices or for potentially new mechanisms not yet applied within IPD. Moreover, we collate the applications with existing blockchain research in the construction industry to indicate their novelty, or if already realized, their alignment.

*Table 3: The 22 identified IPD applications based on the blockchain governance mechanisms. Some were already explored in existing construction blockchain literature.*

| IPD Application of Blockchain Mechanism | Mechanism | Ostrom Principle | Excerpt of Related Blockchain Research in the Construction Industry |
|---|---|---|---|
| M1-1: Scalable management of user identities and rights | M1 | 1a | Current research uses address-based rights as a prerequisite for blockchain applications. None investigate token-based rights. |
| M1-2: Machine participation | M1 | 1a | Robot participation (Lee et al. 2021); Self-owning house (Hunhevicz et al. 2021). |
| M2-1: Representation and ownership of project resources | M2 | 1b | Project bank accounts (Li et al. 2019; Tezel et al. 2021). |
| M2-2: Decentralized funding and investment mechanisms | M2 | 1b | Tokenized investment mechanisms (Tezel et al. 2021; Tian et al. 2020). |
| M3-1: Non-rent seeking and unrestricted matching of project needs with local conditions | M3 | 2a | Reverse auction-based tendering (Tezel et al. 2021); Decentralized design competition (Dounas et al. 2020; Lombardi et al. 2020a). |
| M4-1: Transparent logic for the appropriation and access to resources | M4 | 2b | Financial mechanisms for IPD projects (Elghaish et al. 2020; Rahimian et al. 2021). |
| M4-2: Scalable and self-enforcing shared risk and rewards | M4 | 2b | - |
| M4-3: New incentive structures | M4 | 2b | Token-based incentives for data records (Hunhevicz et al. 2020; Mathews et al. 2017); performance-based life cycle incentives (Hunhevicz et al. 2022b; O'Reilly and Mathews 2019) |
| M5-1: Scaling of collective choices | M5 | 3 | - |
| M5-2: Definition of voting rights for intended power distributions | M5 | 3 | - |
| M6-1: Gamified and scalable sourcing of local actors' knowledge | M6 | 3 | - |
| M7-1: More trust because of transparent user actions and resource flows, as well as predictive automation with smart contracts | M7 | 4a | Blockchain increases trust in supply chains through data tracking, contracting, and transferring resources (Qian and Papadonikolaki 2020). Many papers focus on these aspects. |
| M7-2: Transaction history enables reaction to events and learning from past decisions | M7 | 4a | Many papers focus on triggering financial transactions based on events, e.g. as in (Elghaish et al. 2020; Hamledari and Fischer 2021a). None focus on learning aspect based on transaction history. |
| M8-1: Smart legal contracts | M8 | 4b | Theoretical investigation of intelligent contracts (Mason 2017; McNamara and Sepasgozar 2020, 2021). Performance based smart contracts (Hunhevicz et al. 2022b). |
| M9-1: Reputation tokens for special rights or for credentials | M9 | 4b | - |
| M10-1: Peer-review for project progress, quality, and cost | M10 | 4b | - |
| M11-1: Token-based sanctioning | M11 | 5 | - |
| M11-2: Social sanctioning through transparent action | M11 | 5 | - |
| M12-1: Smart contract based "mini courts" for fast and transparent conflict resolution | M12 | 6 | Blockchain-based dispute resolution platform (Saygili et al. 2021). |
| M12-2: Token-based dispute participation to ensure "skin in the game" | M12 | 6 | - |
| M13-1: Smart contracts ensure that powerful parties cannot solely enforce collective choices and conflict resolution | M13 | 7 | - |
| M14-1: Smart contracts coordinate decision making among organizational tiers | M14 | 8 | - |

### 5.1. M1 – Identity, ownership, and access rights based on addresses and tokens (for OP1a)

IPD projects need clearly defined rights for each actor based on their role in the project (Cheng et al. 2016).

Blockchain allows **scalable management of user identities and rights** (Table 3, M1-1) through address-based identity and/or transferable tokens. Control of access and rights is the foundation for most blockchain applications proposed for the construction industry, as well as all of the other blockchain-based governance mechanisms, such as access to resources (M2-1), access to decentralized markets (M3-1), access to resources (M4-1) as well as shared risk and rewards (M4-2), access to proposal or voting platforms (M5-2), tracking of user and resource actions (M7-1), participation in smart legal contracts (M8-1), peer-review mechanisms (M10-1), jurisdiction systems (M13-1), or coordination among organizational tiers (M14-1). Smart contract logic ensures that only allowed participants can perform certain actions based on their addresses or token-ownership, providing a scalable approach to define user boundaries in IPD multi-party contracts. Research should investigate whether addresses in IPD projects should be controlled at an individual level or by organizational entities. This depends on many aspects, e.g. how profit, liability and risk should be distributed, or whether an incentive system targets individual actors or firms.

Moreover, blockchain only identifies actors through their addresses, therefore allowing for **machine participation** (Table 3, M1-2), e.g. to tender (M3-1) and sign (M8-1) work packages, as well as giving them access to resources (M3-1) and compensating them for their work (M4-2). We see already example of this in research of Lee et al. (2021) where robots get incremental payments for performed work, as well as in the case of no1s1, a self-owning house that can receive funds for provided services, as well as spend funds for maintenance and operations (Hunhevicz et al. 2021). For now, we assume that decision making for IPD will be still human-based, so collective choice mechanisms (M5-1), peer-review mechanisms (M10-1), conflict resolution mechanisms (M12-1), and coordination among organizational tiers (M14-1) does not involve machine participation.

### 5.2. M2 – Tokenization of the resources (for OP1b)

IPD projects require clearly-defined boundaries for the resources, i.e. which specific aspects of project scope and budget are open to all, and which are not (Hall and Bonanomi 2021).

With tokenized project resources, e.g. the project budget, **representation and ownership of project resources** (Table 3, M2-1) can be clearly defined, also allowing monitoring of resources (M7-1). We are so far not aware of any research that proposes tokenization of physical resources in a construction industry context. Some research goes in this direction by exploring how crypto assets can integrate the physical and financial supply chains (Hamledari and Fischer 2021b), or suggesting non-fungible tokens (NFTs) to represent building components (Dounas et al. 2020). Inspiration how and which physical resources to tokenize could also come from the asset-backed tokenization of "Holochain's Commons Engine" or the commodity tokens of the "Economic Space Agency" (Fritsch et al. 2021). Tokenization of the resource would allow to manage digitally one or multiple resource pools with distinct appropriation and payoff functions. Related to the project budget, this relates to the suggestion of blockchain-based project bank accounts for construction projects (Li et al. 2019; Tezel et al. 2021).

In addition, **decentralized funding and investment mechanisms** (Table 3, M2-2) leveraging cryptoeconomics can be explored to extend incentive structures (M4-3). Augmented bonding curves could be one such mechanism to extend current risk/reward structures (M7-1) in IPD by yielding additional profit for invested stakeholders. Having said that, even though tokenized investment mechanisms for construction projects were already proposed (Tezel et al. 2021; Tian et al. 2020), normally a client pays for the project and there is no need to raise funds. Moreover, the power distribution to manage the resources is usually determined by the respective project roles, and not dependent on their point in time when they invest and support the project. Nevertheless, future PDMs might benefit from such new funding and investment mechanisms.

### *5.3. M3 – Decentralized markets to match resources to local needs and conditions (for OP2a)*

Within IPD, key stakeholders often have experience with local conditions, such as availability of labor, material, work routines and other resources. Their early involvement provides the rest of the project team with a holistic understanding of the project conditions (Hall and Bonanomi 2021).

Decentralized markets could improve and extend this with **non-rent seeking and unrestricted matching of project needs with local conditions** (Table 3, M3-1). Projects can find local resources and knowledge important for the success of the project without middleman profiting from facilitating these marketplaces, improving profitability of both the project and contributors. Decentralized marketplaces also allow users of the marketplace, e.g. the IPD stakeholders, project suppliers, and local residents, to collectively define rules. Furthermore, blockchain-based marketplaces can introduce new decentralized market mechanisms, only requiring a blockchain address and/or holding credentials such as reputation tokens. This could lead to more inclusive markets, potentially not only restricted to humans but also machines. We are not aware of any implemented decentralized marketplaces in the construction industry. Along these lines, Tezel et al. (2021) investigated a reverse auction-based tendering mechanism facilitated by smart contracts, and Dounas et al. (2020) and Lombardi et al. (2020b) analyze a decentralized design competition.

### *5.4. M4 – Formalizing appropriation and provision rules with smart contracts (for OP2b)*

In IPD, the risk/reward pool is the main instrument to balance a firm's required participation with the potential reward according to their individual cost structure or accounting practices, their period of involvement in the project and/or their influence on the project's outcome (Cheng et al. 2016).

Smart contracts encode selection criteria and market mechanics visible to everyone and allow to forecast expected behavior according to the formalized rules. **Transparent logic for the appropriation and access to resources** (Table 3, M4-1) can be collectively ensured, especially if the resources are represented in the system through tokens (M2-1). This has been acknowledged for mechanisms of monetary resources in IPD projects (Elghaish et al. 2020; Rahimian et al. 2021). Moreover, smart contracts can ensure **scalable, self-enforcing, and stakeholder specific rules for shared risk and rewards** (Table 3, M4-2), hereby clearly defining provision rules of the system that confirm with the defined appropriation rules. However, this replicates existing IPD allocation rules at the firm level. **New incentive structures** (Table 3, M4-3) for IPDs only possible with blockchain mechanisms can be created. For example, blockchain could be used to issue non-monetary reputation tokens or access-tokens for decentralized markets (M3-1), decision making processes (M5-2), and to ensure "skin in the game" in legal disputes (M12-2). It is also possible to create new ways of token-based rewards (M4-2) and sanctioning (M11-1) at the individual or at the firm level.

Token-based incentives are to date rarely proposed in construction industry literature. Mathews et al. (2017) propose a token to reward parties for maintaining and improving BIM databases. Similarly, Hunhevicz et al. (2020) explore smart contracts and tokens to ensure high-quality data sets in a construction project. Although not token-based, O'Reilly and Mathews (2019) and Hunhevicz et al. (2022b) explored performance-based incentives across life-cycle phases to design and build for the best possible energy performance across phases. Inspiration could come also from outside of the construction industry, e.g. from the Finance 4.0 initiative that explored token-based incentives to address sustainability (Ballandies et al. 2021; Dapp 2019).

### 5.5. M5 – Decentralized proposal and voting platforms (for OP3)

In an IPD context, firms that have signed the multi-party contract are entitled to participate in management group functions and to vote on decisions that concern their work and area of expertise (Ashcraft 2011; Perlberg 2009).

**Scaling of collective choices** (Table 3, M5-1) in IPD could be achieved via decentralized proposal and voting platforms. First, stakeholders can gather opinions and proposals on project spending and execution. Afterwards, they allow for trusted voting on proposals to reach fast decisions even among organizational tiers (M14-1). Finally, they could collectively decide, e.g. through peer-review mechanisms (M10-1), on the appropriation and provision rules of the resources (M4-1) or how to incentivize project stakeholders (M4-3). If the project uses a tokenized resource pool or rewards, approved funds or resources could be automatically released upon approval (M2-1).

Although there are not yet examples in the construction industry, we can find multiple examples of implemented blockchain based decision making mechanisms. *Token Curated Registries* (TCR) (Asgaonkar and Krishnamachari 2018; Wang and Krishnamachari 2019) can be used to manage the validity and functionality of tokens (de la Rouviere 2018). With a TCR users can collectively decide and entries to lists, e.g. to decide on new tokens or changes to existing tokens. Within IPD, a TCR could allow trustworthy and fast collective change processes after the initial project definition to existing tokens or propose new tokens as the IPD participants see fit. Another example for a decentralized governance platform is Politeia[1] for the Decred[2] blockchain, where stakeholders owning the cryptocurrency can upload proposals for network changes and treasury spending and then vote on it. Also, the Aragon project implemented a token based voting platform called Aragon Voice[3]. IPDs could use similar decentralized applications to manage decision making.

Address or token-based access control allows fine-grained **definition of voting rights for intended power distributions** (Table 3, M5-2) among organizational tiers (M14-1), while maintaining scalability of the system. Suited voting forms and decision-making mechanisms would need to be explored in an IPD context. In the blockchain space, various voting mechanisms are proposed that could inspire new ways of voting within IPD. In Decred for example, holders have 1 vote per token (although pooled into larger amounts and locked for an uncertain time period). This approach is anonymous, whereas in 1 vote per person as often used in existing democratic systems, voters need to be identifiable. Another proposed voting mechanism for CPR governance (Dao 2018; Emmett 2019a) includes *quadratic coin lock voting* (Buterin 2016) as a token-based variant of quadratic voting (Weyl and Lalley 2017). The weight of votes is discounted by an exponential function to more prominently value the vote of minority opinions (Fritsch et al. 2021). Finally, in *conviction voting* stakeholders continuously allocate votes in form of tokens to different options that slowly decay if not renewed (Emmett 2019b). This allows to sense user preferences over long time periods and prevent last minute vote swings by large token holders (Fritsch et al. 2021).

### 5.6. M6 – Decentralized prediction markets (for OP3)

To make well informed decisions, decentralized prediction markets could be used for **gamified and scalable sourcing of local actors' knowledge** (Table 3, M6-1), maybe in combination with decentralized markets for local actors (M3-1) and to extend present incentive structures towards external actors (M4-3). Augur[4] is likely the most common implementation of a blockchain based prediction market. To our knowledge there are so far no similar mechanisms within IPD. Nevertheless, research can explore if decentralized prediction markets can be useful in cases where actors are unknown or should remain anonymous, e.g. a betting platform to gauge expected costs of the project.

---

[1] *https://proposals.decred.org/*, accessed 09.12.2021.

[2] *https://decred.org/*, accessed 09.12.2021.

[3] *https://aragon.org/aragon-voice*, accessed 09.12.2021

[4] *https://www.augur.net/*, accessed 09.12.2021.

### 5.7. M7 – Transparent record and automation of transactions (for OP4a)

IPD projects make use of monitoring practices such as "open-book finances" to track financial resources or "Big Room" to collocate stakeholders to commit publicly to work packages and continuously report their progress to the rest of the team (Hall and Bonanomi 2021).

Blockchain creates **more trust because of transparent user actions and resource flows, as well as predictive automation with smart contracts** (Table 3, M7-1). IPD stakeholders can be identified through address-based access control and their transactions tracked visibly to all stakeholders, creating an inherent incentive to behave trustworthy since the other stakeholders can recognize malicious behavior (M11-2). Because of transparent financial transactions, open-book finances is inherently ensured. Furthermore, with resource tokenization implemented, resource flows and appropriation are observable. For example, a blockchain could help monitor the weekly withdrawals of resource units and alert participants to deviations from the cost targets initially estimated with the target value design process. In addition, smart contract automation gives more certainty in the expected transaction logic. Overall, transparency and smart contract automation creates trust in defined explicit incentive structures (M4-3), in collective choices (M5-1), in conflict resolution mechanisms (M12-1), and for coordination among organizational tiers (M14-1). The available **transaction history enables reaction to events and learning from past decisions** (Table 3, M7-2) for the management of user identities and rights (M1-1), ownership of resources (M2-1), decentralized market logic (M3-1), logic for the appropriation and access to resources (M4-1), refining incentive structures (M4-3), the definition of voting rights (M5-2), and the execution of smart legal contracts (M8-1), peer-review mechanisms (M10-1), and decentralized jurisdiction systems (M12-1).

Blockchain increases trust in supply chains through data tracking, contracting, and transferring resources (Qian and Papadonikolaki 2020). Many papers in a construction context focus on transparent and traceable records, e.g. of design data (Erri Pradeep et al. 2021) or construction related quality data (Sheng et al. 2020; Wu et al. 2021). Literature also investigates how to ensure traceability of built asset product information along the construction value chain in various contexts (Kifokeris and Koch 2020; Li et al. 2021; Wang et al. 2020; Watson et al. 2019). Automated and traceable financial transactions are often suggested to enhance financial processes within construction projects (Ahmadisheykhsarmast and Sonmez 2020; Chong and Diamantopoulos 2020; Das et al. 2020; Elghaish et al. 2020; Di Giuda et al. 2020; Hamledari and Fischer 2021a; Nanayakkara et al. 2021; Ye and König 2021).

### 5.8. M8 – Digital signatures for tamper-proof commitments (for OP4b)

To improve accountability within IPD, stakeholders can commit to work packages by signing a blockchain transaction. This enables **smart legal contracts** (Table 3, M8-1) when linked to terms encoded in smart contracts for trackable and automatic execution. Construction literature theoretically suggested smart legal contracts (Maciel and Garbutt 2020; Shojaei et al. 2020), also termed intelligent contracts (Mason 2017; McNamara and Sepasgozar 2020, 2021). Based on agreed terms about committed work or performance data (Hunhevicz et al. 2022b), progress and completion can be tracked and confirmed (M7) to ensure accountability. Because machines can hold access rights, they also could commit to work packages and participate in smart legal contracts.

### 5.9. M9 – Reputation tokens (for OP4b)

The concept of **reputation tokens for special rights or for credentials** (Table 3, M9-1) presents an interesting approach to reward or punish IPD participants (M4-3). Instead of monetary incentives, reputation tokens based on stakeholder accountability could give access to extended governance functions (M5-2) or could be used for credentials in decentralized markets for later projects (M3-1).

### 5.10. M10 – Decentralized peer-review mechanisms (for OP4b)

Decentralized **peer-review for project progress, quality, and cost** (Table 3, M10-1) can be implemented with blockchain. For example, the Covee protocol realized a smart contract based peer review mechanism to determine fair profit distribution for decentralized collaborative teams (Dietsch et al. 2018). Anonymous work contributors get rewarded with cryptocurreny according to their peer-review score. A combination of blockchain-signed work packages, reputation tokens, and decentralized peer-review mechanisms could create a digital "big room platform" to ensure presence and accountability within IPD, and to evaluate appropriate rewards (M4-3) and sanctions (M11-1).

### 5.11. M11 – Transparent and self-enforcing sanctions (for OP5)

Graduated sanctions are often not explicitly implemented in IPDs (Hall and Bonanomi 2021). At most, the weekly public reporting of "Planned Percent Complete" (PPC) (Thomsen et al. 2009) acts as an early stage of social sanctioning (Kenig et al. 2010). In the case of continuous non-conformance or underperformance, the removal of individual participants and/or firms can be necessary (Cheng et al. 2016).

Blockchain can be an opportunity to reimagine and improve upon graduated sanctioning for IPD projects through decentralized and self-enforcing **token-based sanctioning** (Table 3, M11-1), e.g. loss of access tokens, loss of reputation tokens, or decrease in value of monetary tokens. Underperformance is also visible to everyone leading to **social sanctioning through transparent action** (Table 3, M11-2). In many cases this might be enough to ensure accountability but could be gradually combined with loss of tokens e.g. for access to the financial project rewards or even the project itself.

### 5.12. M12 – Decentralized jurisdiction systems (for OP6)

IPD projects craft conflict resolution mechanisms such as project decision protocols (Ashcraft 2011) or liability waivers (Sive and Hays 2009) that include clear dispute resolution strategies intended to avoid costly litigation proceedings.

Blockchain enables **smart contract based "mini-courts" for fast and transparent conflict resolution** (Table 3, M12-1) in IPDs. An exemplary decentralized jurisdiction system is already implemented in the *Aragon Court*[5] to resolve subjective disputes that cannot be resolved by smart contracts alone. A global network of guardians helps intervene and arbitrate disputes. Saygili et al. (2021) already propose a decentralized blockchain-based online dispute resolution platform to resolve construction disputes. To incentivize compliance and accountability in a decentralized jurisdiction, **token-based dispute participation to ensure "skin in the game"** (Table 3, M12-2) is suggested. In case of non-compliance with the rules or the verdict, tokens at stake can be sanctioned.

### 5.13. M13 – Ensure decisions are made by affected parties (for OP7)

In IPD projects, project sponsors trade decision-making autonomy for consensus mechanisms among project team members (Hall and Bonanomi 2021). In other words, authority is given by the project owner to the project participants to self-organize and self-govern the project. Blockchain transparency and censorship resistance enables **smart contracts to ensure that powerful parties cannot solely enforce collective choice and conflict resolution** (Table 3, M13-1) in IPDs (M5-2, M12-1, and M14-1). Ideally, decisions should be only possible to be made and challenged by actors that are also affected.

---

[5] *https://court.aragon.org/, accessed 09.12.2021.*

*5.14.   M14 – Coordination rules across nested enterprises (for OP8)*

Large IPD projects have multiple nested management levels, including senior management team for executive leadership, a cross-functional project management team to coordinate project management activities, and functional teams that handle the direct work execution and organization (Ashcraft 2011; Laurent and Leicht 2019). Therefore, it will be important that **smart contracts coordinate decision making among organizational tiers** (Table 3, M14-1). This is either according to existing nested management levels of IPD or for new forms of organization better suited for fast information propagation and reactions to local events enabled by blockchain-based project governance.

**6.   IPD on the Crypto Commons: An Overview of the Conceptualization**

Figure 2 summarizes the overall conceptualization, visualizing the OPs (Table 1), the 14 identified blockchain governance mechanisms (Table 2), and the 22 proposed applications for IPDs (Table 3).

The interaction arrows indicate then graphically the described connections in Section 5 between the different IPD blockchain applications. The arrows are either outgoing, meaning they are a prerequisite or support other applications, or incoming, meaning they are supported or enabled by other applications. The width of the arrows and/or number of incoming connections can give an approximate indication of the importance of applications (outgoing) or prerequisites to build applications (incoming) within the overall conceptualization.

Applications that stand out regarding important prerequisite for IPD on the crypto commons are: M1-1 defining boundaries for the users through addresses and tokens, and M7-2 monitoring the presence of users and resources for fast system reaction and learning based on transparent record of automation and transactions. Applications that depend on many interactions of other applications in the system are in decreasing order of incoming connections: M4-3 for new incentive structures to influence the system participants towards collective action, M3-1 decentralized market structures to match project needs with local conditions, M5-2 for definition of voting rights to create intended power distributions, and M14-1 coordination among organizational tiers.

It is likely that not all connections have been identified and the interactions need to be updated when more research investigates individual applications and/or the interaction between them.

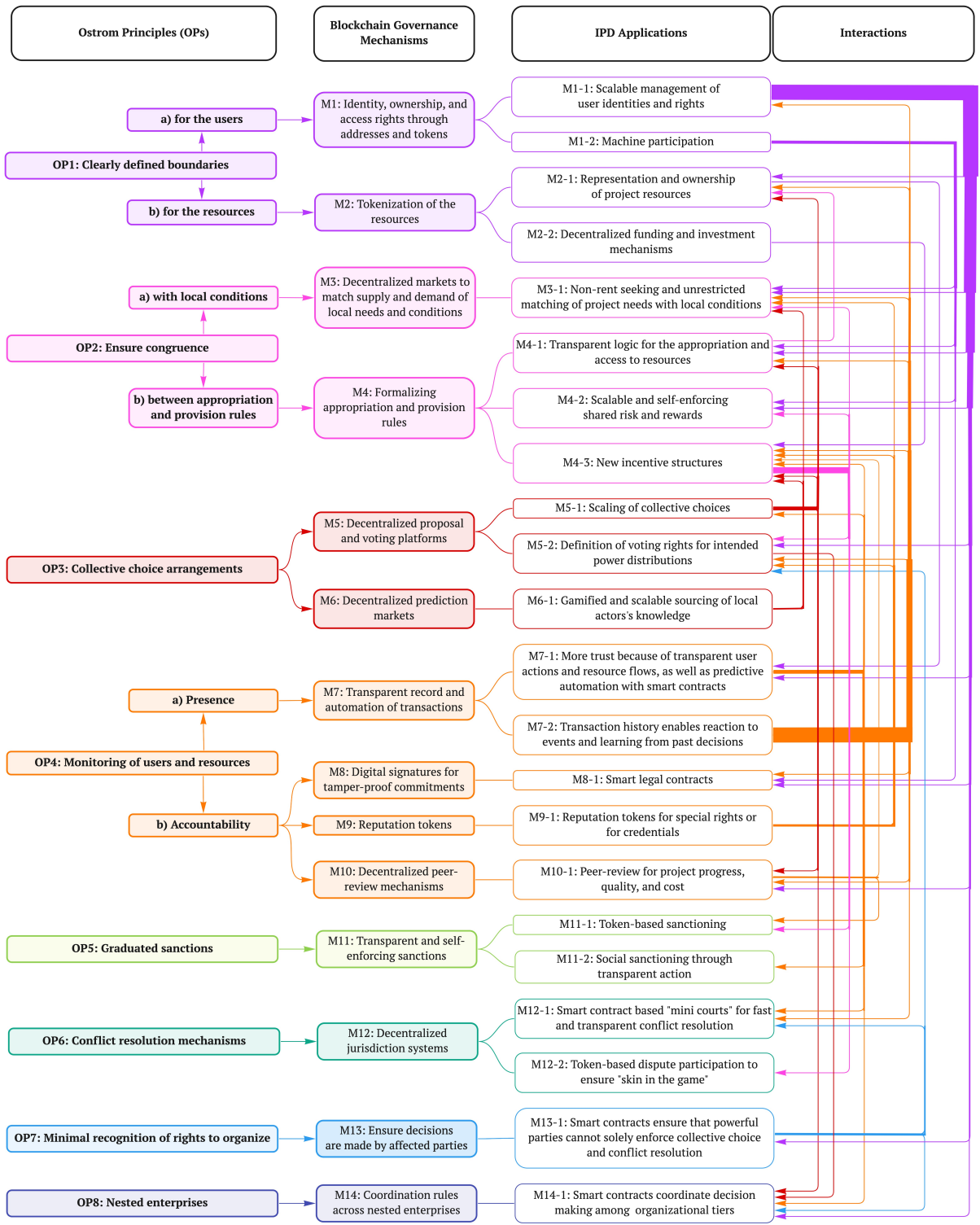

*Figure 2: Overview of the conceptualization of IPD on the crypto commons: the 12 blockchain governance mechanisms for the eight OPs as identified in Section 4, and the 22 identified applications for IPD and their interactions outlined in Section 5.*

## 7. Discussion

### 7.1. Impact

The novel proposition of this paper is to create blockchain-based governance structures for IPD construction projects using the OPs as design guidelines. Trusted digital processes together with cryptoeconomic incentive mechanisms can align stakeholders, both human and machine, to better collaborate towards the overall project success. We see two main scenarios for the application of blockchain-based governance mechanisms as introduced in this paper.

First, blockchain based governance could address tradeoffs of current relational contracting approaches to improve IPDs. Relational contracting is well-suited to deal with contractual hazards of "displaced agency" (Henisz et al. 2012) found in the fragmented (Fergusson and Teicholz 1996; Howard et al. 1989; Levitt and Sheffer 2011) and loosely coupled (Dubois and Gadde 2002a) construction project structures. However, relational contracts also comes at various costs (Henisz et al. 2012) that could be improved through blockchain-based governance mechanisms, e.g. reduced competition with scalable decentralized market structures, or lengthier processes for decision-making with decentralized decision-making platforms.

Second, the introduced blockchain mechanisms could be used to build new forms of project delivery coordinated on the crypto commons. Construction projects can be characterized by complexity (Bertelsen 2003; Dubois and Gadde 2002a; Gidado 1996). Research suggests that bottom-up management and self-organization are better suited than hierarchical approaches to manage complexity (Bertelsen and Koskela 2004; Helbing and Lämmer 2008). The OPs introduce guidelines to achieve this for CPR scenarios. Since IPD can be described as a CPR scenario, blockchain governance mechanisms could improve IPD-like project deliveries by creating better bottom-up and self-organizing project structures, while still allowing for scalable coordination mechanisms.

This is aligned with the emerging organizational form of a decentralized autonomous organization (DAO), which is a blockchain-based system that enables people to coordinate and govern themselves mediated by a set of self-executing rules deployed on a public blockchain, and whose governance is decentralized (Hassan and De Filippi 2021).

Sreckovic and Windsperger (2020) already proposed the evolution of the construction industry organization towards DAOs. Lombardi et al. (2020b) and Dounas et al. (2020) even prototyped a DAO for decentralized coordination of the design finding process through smart contracts. Their research suggests that construction project governance as a DAO is at least in a prototyping context technically feasible. Other ongoing research explores how decision making of a self-owning house can be coordinated through a DAO (Hunhevicz et al. 2021).

Emerging examples of DAO frameworks (Faqir-Rhazoui et al. 2021) resemble many of the identified governance mechanisms. The introduced research directions based on Ostrom's design principles might help to design governance building blocks towards project delivery coordinated through DAOs.

### *7.2. Design Challenges*

Designing new blockchain based governance systems is challenging. Cila et al. (2020) identified six design challenges (see Table 4) that are further discussed below in the context of IPDs on the crypto commons.

#### 7.2.1. Tracking

While transparent monitoring is essential to manage commons, it could lead to privacy concerns regarding the community-based data (Cila et al. 2020). Especially in public blockchain systems, traditional data privacy solutions are hard or even impossible to implement. It needs to be carefully evaluated what data needs to be transparently stored to enable IPD governance, how construction stakeholders perceive implications of sharing this data, and potential measures to maintain a suited level of privacy without hindering the monitoring.

#### 7.2.2. Coding

A major challenge is decide, represent, and encode values in artificial commons (Cila et al. 2020). It tends to be easier to focus on economic and quantifiable values in a blockchain system (e.g. monetary values through market pricing mechanisms) than social and qualified values (e.g. reputation mechanisms). However, in commons non-monetary values often play an important role (Fritsch et al. 2021). Also, in IPDs, both quantitative and informal systems are used. Future research should investigate value flows in IPD, as well as how to encode them into tokens and incentive systems without suppressing creativity and teamwork with too rigid and inflexible smart contract structures.

Furthermore, incentives give people a sense of agency, yet at the same time they can have downsides of forced conformity with collectively set rules (Cila et al. 2020). At some point earning rewards might become a duty to not be excluded from the system and rewards cause efforts to shift towards the actions that will be rewarded, causing potentially unforeseen negative secondary effects (Cila et al. 2020). This needs to be subject of further study when designing project delivery mechanisms.

Artificial commons needs to find a balance between so trading individual gains for the greater good of the community (Cila et al. 2020). IPDs have differences to natural commons that need to be considered. For example, resources in IPDs are consumed intentionally over time, whereas in natural commons they are renewable. Natural commons also have an infinite lifetime (if the community is able to sustain them), while IPDs only last for the duration of a project (although this could be several years). And the product is owned by the project sponsor, whereas natural commons are not owned by any of the involved parties.

Overall, there is a need for more research to thoroughly understand current and new IPD mechanisms, how they contribute to the success of IPDs, and how they need to be set up in different project settings. Methods used to design and test such mechanisms should be able to reflect the complex nature of construction projects (Bertelsen 2003; Dubois and Gadde 2002a; Gidado 1996). Previous research used game theory for the evaluation of profit distribution (Teng et al. 2019) and target value design (Jung et al. 2012), agent based simulations to assess the evolution of collaboration (Son and Rojas 2011), or mechanism design to investigate new incentive structures (Han et al. 2019).

### 7.2.3. Negotiation

The last dilemma concerns how to preserve human reasoning and debate in a system of formalized and algorithmic logic (Cila et al. 2020). Also for IPD, there are major risks involved in ex-ante designs of smart contract, where system engineers need to account from the beginning for all expected cases. Therefore, the governance system should be able to adjust over time to exceptions and design errors through community input. But even with such governance processes embedded, the process will likely only start after the first failure already happened. A stepwise and careful adaption with extensive testing of these systems will be desirable.

*Table 4: Design challenges for crypto commons (Based on the identified blockchain governance design challenges by Cila et al. (2020)).*

| Design Challenge | Type | Description |
|---|---|---|
| Transparency vs. Privacy | Tracking | Crypto Commons must be monitored, but transparent tracking could lead to privacy concerns. |
| Economic vs. Social Values Quantified vs. Qualified Values | Coding | Values of the Crypto Commons must be encoded in a representative way. This can be especially challenging for social or qualified values. |
| Incentivisation vs. Manipulation | Coding | Crypto Commons must encode incentives without causing unjustified manipulation and exclusion of stakeholders. |
| Private vs. Collective Interests | Coding | Encoding rules for Crypto Commons must weigh individual gains of stakeholders against the greater good of the community. |
| Human vs. Algorithmic Governance | Negotiation | Crypto commons must preserve human reasoning and debate in a system of formalized and algorithmic logic. |

*7.3. Challenges related to the construction industry context*

While technical and system design challenges towards a blockchain governed project delivery can be proactively addressed, there are many inherent construction industry barriers. Other scholars have already investigated barriers and socio-technical challenges for blockchain in the construction industry (Li et al. 2019). The provided frameworks likely apply also for the proposed system. Below we highlight some of the key challenges.

The level of digitalization in the construction industry is still low (Agarwal et al. 2016; Barbosa et al. 2017). Blockchain-governed PDM require an extensive digital base-line of project related data. As long as this data is not available, the proposed mechanisms cannot make use of it to govern construction projects.

Moreover, the fragmented construction industry structure poses major challenges in the adoption of systemic innovations (Hall et al. 2018), e.g. as in the case of BIM (Papadonikolaki 2018). Blockchain based governance for construction PDMs likely falls into the same category of systemic innovations, since value of the solution only comes at scale. At the same time, cryptoeconomic governance promises to reduce implications of fragmentation through incentives across phases and trades (Hunhevicz et al. 2022a). Nevertheless, it will be challenging to organically grow adoption.

Finally, there are major legal implications with such new solutions. Research needs to investigate how smart contract code can conform with law and regulations (De Filippi and Hassan 2016).

*7.4. Limitations*

For blockchain based governance processes, the underlying blockchain infrastructure is a key component to success. For simplicity, this paper only refers to blockchain, but there are many kinds of DLTs suitable for different types of use cases. The choice of the right type is not part of this work, but should be considered once a use case will be implemented (Hunhevicz and Hall 2020). Many parameters such as security, throughput, privacy, approaches to smart contracts, and others need to be assessed. The still early technological state and the many different available solutions (Ballandies et al. 2021; Spychiger et al. 2021b) make this challenging. The paper assesses blockchain for CPR and IPD based on the assumption of public permissionless blockchains, since they align with the promise to govern decentralized economic coordination. When using other DLT options, the affected properties need to be adjusted. Future research should assess suited technical infrastructure to realize the proposed blockchain governance mechanisms in an IPD context.

Furthermore, the paper acts only as a starting point to conceptualize the connections between blockchain, CPR theory, and IPD. As a next step, further research is needed to validate the conceptualization. This includes validation of both the individual mechanisms and applications as well as their interaction. The contribution of this paper is limited as a proposed conceptualization, meant to underpin future research efforts that can validate and extend the conceptualization.

## 8. Conclusion

The paper extends the thinking around blockchain as an institutional innovation for the delivery of construction projects. It exploits the theoretical connection between both blockchain and IPDs as a CPR scenario to offer a systematic starting point how blockchain can support and evolve PDMs by creating novel governance mechanisms.

For that the paper introduces a conceptualization of blockchain-based governance applications for IPDs on the crypto commons. Twenty-two applications for IPD were identified based on fourteen mechanisms of blockchain for the governance of CPR scenarios proposed to encode the eight OPs. The conceptualization is useful to think more structured and modular about blockchain building blocks to govern construction projects collectively on the "crypto commons". Furthermore, the conceptualization can support the thinking around how blockchain could improve current IPD concepts, potentially lead to the next generation of PDMs, or ultimately end in novel project coordination through DAOs. On the one hand, blockchain-based governance mechanisms promise to facilitate trusted, scalable and efficient bottom-up coordination mechanisms that cope with complexity and displaced agency in construction projects. On the other hand, blockchain-based project delivery offers exciting new opportunities for machine participation.

Even though the paper introduced a coherent conceptualization for blockchain-based governance of PDMs, it requires further validation through proof of concepts investigating the feasibility of individual and combined mechanisms. For that the paper discusses challenges related to the early state of blockchain technology, the difficulties in designing blockchain-based governance systems, and the industry-related challenges to overcome.

The paper primarily targets the construction industry, but the identified blockchain governance mechanisms and applications could eventually be transferred to other cases of real-world commons.

## 9. Funding

This research did not receive any specific grant from funding agencies in the public, commercial, or not-for-profit sectors.

*2012*, American Society of Civil Engineers, Reston, VA, 556–563.

Kenig, M., Allison, M., Black, B., Burdi, L., Colella, C., Davis, H., and Williams, M. (2010). *Integrated Project Delivery for Public and Private Owners*. National Association of State Facilities Administrators, Construction Owners Association of America, The Association of Higher Education Facilities Officers, Associated General Contractors of America and The American Institute of Architects.

Kifokeris, D., and Koch, C. (2020). "A conceptual digital business model for construction logistics consultants, featuring a sociomaterial blockchain solution for integrated economic, material and information flows." *Journal of Information Technology in Construction*, International Council for Research and Innovation in Building and Construction, 25(29), 500–521.

Kim, K., Lee, G., and Kim, S. (2020). "A Study on the Application of Blockchain Technology in the Construction Industry." *KSCE Journal of Civil Engineering*, Springer Verlag, 24(9), 2561–2571.

Kovács, G., and Spens, K. M. (2005). "Abductive reasoning in logistics research." *International Journal of Physical Distribution and Logistics Management*, Emerald Group Publishing Limited.

de la Rouviere, S. (2018). "Saving The Planet: Making It Profitable To Protect The Commons." <https://medium.com/@simondlr/saving-the-planet-making-it-profitable-to-protect-the-commons-50393906fe22>.

Lahdenperä, P. (2012). "Making sense of the multi-party contractual arrangements of project partnering, project alliancing and integrated project delivery." *Construction Management and Economics*, 30(1), 57–79.

Laurent, J., and Leicht, R. M. (2019). "Practices for Designing Cross-Functional Teams for Integrated Project Delivery." *Journal of Construction Engineering and Management*, 145(3), 05019001.

Lee, D., Lee, S. H., Masoud, N., Krishnan, M. S., and Li, V. C. (2021). "Integrated digital twin and blockchain framework to support accountable information sharing in construction projects." *Automation in Construction*, Elsevier, 127, 103688.

Levitt, R. E. (2011). "Towards project management 2.0." *Engineering Project Organization Journal*, Engineering Project Organization Society, 1(3), 197–210.

Levitt, R. E., and Sheffer, D. A. (2011). "Innovation in Modular Industries: Implementing Energy-Efficient Innovations in {US} Buildings1." *Proceedings of Conference on Energy, Organizations and Society, UC Davis*, 1–29.

Li, J., Greenwood, D., and Kassem, M. (2019). "Blockchain in the built environment and construction industry: A systematic review, conceptual models and practical use cases." *Automation in Construction*, Elsevier, 102, 288–307.

Li, J., and Kassem, M. (2021). "Applications of distributed ledger technology (DLT) and Blockchain-enabled smart contracts in construction." *Automation in Construction*, Elsevier B.V.

Li, X., Wu, L., Zhao, R., Lu, W., and Xue, F. (2021). "Two-layer Adaptive Blockchain-based Supervision model for off-site modular housing production." *Computers in Industry*, Elsevier B.V., 128, 103437.

Lombardi, D., Dounas, T., Cheung, L. H., and Jabi, W. (2020a). "Blockchain Grammars for Validating the Design Process." *Blucher Design Proceedings*, Editora Blucher, São Paulo, 406–411.

Lombardi, D., Dounas, T., Cheung, L. H., and Jabi, W. (2020b). "Blockchain Grammars for Validating the Design Process." *Blucher Design Proceedings*, Editora Blucher, São Paulo, 406–411.

Luo, L., He, Q., Jaselskis, E. J., and Xie, J. (2017). "Construction Project Complexity: Research Trends and Implications." *Journal of Construction Engineering and Management*, American Society of Civil Engineers (ASCE), 143(7), 04017019.

Machart, F., and Samadi, J. (2020). *The state of blockchain governance*.